\documentstyle[12pt,epsf]{article}
\newcommand{\beq}{\begin{equation}}
\newcommand{\eeq}{\end{equation}}
\newcommand{\bea}{\begin{eqnarray}}
\newcommand{\eea}{\end{eqnarray}}
\newcommand{ \nn}{\nonumber}

\newcommand{\gtrsim}{\ \rlap{\raise 2pt\hbox{$>$}}{\lower 2pt
\hbox{$\sim$}}\ }
\newcommand{\lessim}{\ \rlap{\raise 2pt\hbox{$<$}}{\lower 2pt
\hbox{$\sim$}}\ }

\newcommand{\np}[1]{Nucl. Phys. {\bf #1}}
\newcommand{\pl}[1]{Phys. Lett. {\bf #1}}
\newcommand{\pr}[1]{Phys. Rev. {\bf #1}}

\makeatletter
\if@twoside
\else \oddsidemargin 00pt \evensidemargin 00pt
\fi
\let\@eqnsel = \hfil

\expandafter\ifx\csname mathrm\endcsname\relax\def\mathrm#1{{\rm
#1}}\fi
\@ifundefined{mathrm}{\def\mathrm#1{{\rm #1}}}{\relax}

\makeatother

\begin{document}
\thispagestyle{empty}
\null
\hfill FTUV/96-28,IFIC/96-32

\hfill hep-ph/9606218

\vskip 1.5cm

\begin{center}
{\Large \bf      %INSERT TITLE
 LEFT-HANDED NEUTRINO DISAPPEARANCE
\par  \vspace{.5cm}
PROBE OF NEUTRINO MASS AND
\par \vspace{.5cm}
CHARACTER } \vskip 2.em
{\large		%INSERT NAMES
{\sc G. Barenboim$^{1,2}$, J. Bernab\'eu$^1$  and O. Vives$^{1,2}$
}  \\[1ex] %INSERT ADDRESS
{\it $^1$ Departament de F\'\i sica Te\`orica, Universitat
de Val\`encia}\\
{\it $^2$ IFIC, Centre Mixte Universitat
de Val\`encia - CSIC} \\
{\it E-46100 Burjassot, Valencia, Spain} \\[1ex]
\vskip 0.5em
\par}
\end{center} \par
\vfil
{\bf Abstract} \par
We explore the sensitivity to a non vanishing
neutrino mass offered by dynamical observables, i.e., branching
ratios and polarizations. The longitudinal polarization
in the C.M. frame decreases
by a 4\% for $D^+ \rightarrow \tau^+ \nu_\tau$ and $m_{\nu_\tau}=24$
MeV.
Taking advantage of the fact that the polarization is
a Lorentz variant quantity, we study the polarization effects in
a boosted frame.
By means of a neutrino beam, produced by a high velocity boosted
parent
able to flip the neutrino helicity, we find that an enhanced
left-handed
neutrino deficit, induced by a Wigner rotation, appears.
\par
\vskip 0.5cm
\noindent May 1996 \par
\null
\setcounter{page}{0}
\clearpage

The most direct searches for neutrino mass are the experiments
which seek a kinematical consequence of this mass in some physical
processes \cite{uno}.
Nevertheless, for massive neutrinos there are noticeable dynamical
effects that affect the predictions for physical observables, like
transition probabilities and polarizations \cite{unob}.
In this work we explore
the sensitivity to a non-vanishing  neutrino mass offered by
observables which are modified by these two facts:

a) when $m_\nu \neq 0$ definite chirality in the weak interaction
does not mean definite helicity;

b) when both $m_\nu \neq 0$ and the presence of a boost, helicity
is not Lorent invariant and the transformed state of an helicity
state is a linear combination of different helicity states.

The second fact implies, in particular, that Lorentz variant
observables, like the components of polarization, through the
so-called
``rotation of the spin in a Lorentz transformation" \cite{dos},
should be modified from the massless case result by enhancement
factors associated to the boost momentum.
These ideas are applied  here to the pure-leptonic decays of
pseudoscalar mesons $P \longrightarrow l  \; \nu_l$.
The dynamical mass-effects are analysed in the rates and the
components
of polarization in the final state.

Upper bounds on the $m_{\nu_\mu}$, the mass of $\nu_\mu$, come from
studies of the decay $\pi \longrightarrow \mu  \; \nu_\mu$ using
pions
at rest or ones in flight, for which the measurement of the muon
momentum \cite{tres}, combined with information on the pion and
muon masses, provides the information. The most stringent existing
bound
is $m_{\nu_\mu} \leq 0.16$ MeV at 90$\%$ C.L..
The polarization of the muon from pion decay is
determined from the relative
electron rate at the momentum end point in a direction
opposite to the $\mu$-spin in $\mu$-decay. This  has been  used
\cite{cuatro}
to obtain the neutrino helicity $ \mid h_{\nu_\mu} \mid \geq 0.9968$,
at a 90$\%$ C.L..

Limits on $m_{\nu_\tau}$, the mass of $\nu_\tau$, come from studies
of the invariant mass and energy of the hadronic system in
$\tau \longrightarrow \nu_\tau + \mbox{hadrons}$. The bigger the
invariant mass is, the less energy is available to make the mass
of $\nu_\tau$ in the final state. The most stringent laboratory bound
is  \cite{cinco} $m_{\nu_\tau} \le 24$ MeV at 95$\%$ C.L. .
Recent analyses \cite{seis} show
that a tau-neutrino of mass between 1 MeV and 25 MeV can have a host
of interesting astrophysical and cosmological consequences.
In light of this interest, we consider
the $\nu_\tau$-mass effect on the
pure-leptonic decay of charmed mesons $D \longrightarrow \tau
\; \nu_\tau$ and
$D_s \longrightarrow \tau \; \nu_\tau$. The pure-leptonic decays of
charm are
showing up \cite{siete} recently and their evidence and precision
will improve  over the next few years as large data samples become
available. The present knowledge can be summarized \cite{ocho}
in the upper limit on the $D^+$ decay, $B ( D^+ \longrightarrow \mu^+
\nu_\mu ) \le 7.2 \cdot 10^{-4}$, at a 90$\%$ C.L., and
the evidence on the $D_s^+$ decay, $B ( D_s^+ \longrightarrow \mu^+
\nu_\mu ) = ( 5.9 \pm 2.2 ) \cdot 10^{-3}$.
The $\tau^+ \; \nu_\tau$ channels are a factor 2.6 and 9.8 more
probable, respectively, but their search is difficult due to the
presence of at least two neutrinos in the final state.

For the pure-leptonic decay of the pseudoscalar meson
$P^+ \longrightarrow l^+ +  \, \nu_l$, the decay amplitude is
given by
\bea
T_{\lambda, \lambda^\prime } =
\langle l^+ (k_l , \lambda^\prime ) \; \nu_l (k_\nu,\lambda )
\mid T \mid P^+ (q) \rangle =
i \; \frac{G_F}{\sqrt{2}} f_P \; q_\mu \;
\bar{u}_{\nu_l}(k_\nu,\lambda )
\gamma^\mu (1- \gamma_5 ) \, v_{l}(k_l,\lambda^\prime )\nn
\eea
where $f_P$ is the $P$-meson decay coupling constant.

For $m_\nu = 0$ the left-handed chirality of charged current weak
interactions then leads to a definite helicity -1 for the outgoing
neutrino and, a fortiori, for the $l^+$ in the C.M. frame.
However,
for $m_\nu \neq 0 $ the helicity has no definite value and we talk
about the polarization components of the neutrino.
The transition probability for a given helicity $\lambda$ of the
neutrino is $ \Gamma_\lambda  =  \mid \vec{k} \mid
\; \mid T_\lambda \mid^2 / (8 \pi \; m_{P}^2 )$,
where $m_P$ is the meson mass and
the squared amplitude, $\mid T_\lambda \mid^2 =
\sum_{\lambda^\prime}
\mid T_{\lambda \lambda^\prime} \mid^2$, is given by
\bea
\mid T_\lambda \mid^2 \; = \; 2 G_F^2 f_P^2
\{ 2(k_\nu.q)(k_l. q) \, -\, (k_\nu.k_l )q^2 \,
-\,2 \lambda \, m_\nu(k_\l .q)(s.q) \,  + \,  \lambda\, m_\nu
(k_l.s)q^2 \}
\label{pol}
\eea
The momentum $\mid \vec{k} \mid$ in C.M., determined by
kinematics, is related to $m_\nu$, $m_\nu^2 = m_P^2 + m_l^2 - 2 m_P
\sqrt{ m_l^2 +
\mid \vec{k}\mid^2} $. This method  is currently
being used for the determination of $m_{\nu_\mu}$ in
 the $\pi$-decay case. The polarization four-vector
$s^\mu$  in Eq (\ref{pol}) satisfies $
s.s \; = \; -1 $  and $s.k_\nu\;=\; 0 $
and its choice is dictated by the polarization basis.
For $m_\nu \neq 0$, in the rest frame of the neutrino,
$s^\mu = ( 0, \vec{s} )$ with $\vec{s}$ the unit vector along the
chosen direction of
quantization.

The neutrino longitudinal polarization can be easily calculated
putting  $\vec{s}$ in the direction
of motion of the neutrino, yielding
\bea
\label{cmpol}
P_{long} \; = \;  \frac{(E-W)\mid \vec{k} \mid }{(WE -\mid
\vec{k}\mid^2)}
\label{long}
\eea
where $E$ and $W$
are the energies of the neutrino and charged lepton, respectively.

To obtain the decay rate is staightforward:
the sum over the helicity $ \lambda$ gives
\bea
 \Gamma \; = \; \sum_\lambda \Gamma_\lambda \; = \; \frac{G_F^2
f_P^2}
{8\pi m_P^2 }
\; \mid \vec{k} \mid \,
 \left(  m_P^2 \left( m_l + m_{\nu} \right)^2
- \left( m_l^2 - m_{\nu}^2 \right)^2
 \right) \nn
 \eea
Having this, we can easily calculate the ratio between the decay
rates of
a meson to two different leptons, namely, tau lepton and muon,
getting
an answer independent of $f_P$.

So far we have seen that dynamical effects of a neutrino mass
different from
zero show up in the leptonic decay of a meson. This effect
depends on the mass of the decaying meson as well as
on those of the daughter-products. Therefore, situations exist
where these effects become more important, and our aim now is to
look for them.

We are interested in a longitudinal polarization in C.M. different
from -1,
it is easy to see that the most favourable situation
takes place when little
phase space is left.
Having this in mind,
we must search for a decay where the mass of the charged
lepton and
the mass of the meson are as close as possible.

For the tau neutrino there is
a very favourable situation with the $D$ meson. Another possibility
is to use the decay of
the $D_s$, in this case more phase space
 and more $\nu$s are available, although the neutrino mass effects
are somewhat diluted. For the muon neutrino the best decay for these
purposes
 is the decay of the pion.

According to what was stated before,
the decay $D^+ \longrightarrow \tau^+ + \nu_\tau $
is the best decay
to look for dynamical effects of the tau-neutrino mass. This decay
has not
been observed yet, and we can expect to
see it in the future in a tau-charm factory \cite{taucharm}.
 Its measurement is of
interest to determine $f_D$. For our purposes to study the neutrino
mass
effects,
we calculate the ratio between the decays
$D^+ \longrightarrow \tau + \nu_\tau$ and $D^+ \longrightarrow \mu +
\nu_\mu$
for different masses of the tau-neutrino. The result is given in
Table I,
showing a 3$\%$ effect at most for $m_{\nu_\tau}$ up to 24 MeV.
For the longitudinal polarization of the neutrino in the C.M.
we can see from
Eq. (\ref{long})
how different from -1 are  the results
in terms of the
neutrino mass. In the best situation (the heaviest possible
neutrino compatible with the experimental limit)
the value of $-P_{long}$ differs from one in a 4$\%$.
So one should have a good experimental precision
for the effect to show up. This is envisageable in a tau-charm
factory, measuring the hadron spectrum  and angular distribution
from tau decay
coming from a $D$, even if we cannot reconstruct the tau-direction
\cite{ref}.
Besides that a tau-charm factory has the capability of tightly
controlling
the background  and systematic errors \cite{taucharm}.

On the other hand, the $D_s$ decay may be used too as a tool
to get  experimental evidence of neutrino mass.
This decay leaves a bigger
phase space, shrinking the numerical predictions as
can be seen in Table II: 1$\%$ effect at most for both branching
ratio and polarization.
The statistics is however better.

We turn our attention to the muon-neutrino.
The  most attractive scenario to look for a hint of $m_{\nu_\mu}$
different from zero is pion decay.
 This decay has been widely observed and the precision
in the muon momenta is  strikingly high.
We have explored the possibility of using the
decay rate and the longitudinal polarization to look
for muon-neutrino mass effect. They cannot compete
with the kinematical determination.

This analysis shows that the mass effects based on the
dynamical observables could be of interest for the
$\nu_\tau$. Even more,
although the branching ratio is Lorentz invariant,
the longitudinal polarization  {\bf{is not}}. The helicity is
a Lorentz invariant quantity
only for massless particles. In this way, if the
neutrino has zero mass, it will have negative helicity in any frame.
On the other hand, if we allow the neutrino to have a mass, its speed
is no longer  1, so that boosting  appropiately the reference
frame, i.e. the decaying meson, we can easily generate the opposite
helicity \cite{cm}.
The rest of the paper is going to be  devoted to analise
neutrino-mass effects in a boosted system.

The effect of a Lorentz transformation on the polarization
appears as a rotation of the
polarization  three-vector relative to the particle
three-momentum:
 this is called a Wigner rotation \cite{dos}.
 Then we can see \cite{martin} that the
amplitudes for a process $ A( q )\rightarrow a ( k_a,\lambda_a ) b (
k_b,
\lambda_b )$ transform in the following way ,
\bea
T_{\lambda_a \lambda_b}(q,k_a) \; = \; \sum_{\lambda_a^{\prime}
\lambda_b^{\prime}}
(-)^{\lambda_b^{\prime} - \lambda_b}\;
d^{1/2}_{\lambda_a^{\prime} \lambda_a} (\omega_a)\;
d^{1/2}_{\lambda_b^{\prime} \lambda_b}
(\omega_b)\;T_{\lambda_a^\prime
\lambda_b^\prime}(q^{\prime},k_a^{\prime})  \nonumber
\eea
where $\omega_a$ and $\omega_b$ are the Wigner rotations for the
particles $a$
and $b$ under a boost which transforms their momenta :
$k_i^{\prime}=\Lambda
k_i$ .
The rotations are as follows,
\bea
\sin \omega_i = \frac {m_i \sinh \kappa \sin
\theta^\prime_i}{\mid\vec{k}_i
\mid}
\;\;\;\; , \;\; \;\;
\cos \omega_i = \frac {\mid \vec{k}_i^\prime \mid \cosh \kappa -
E_i^\prime
\sinh \kappa \cos \theta^\prime_i}{\mid \vec{k}_i \mid}
\label{wignercos}
\eea
where $\kappa$ is the parameter of the boost related to the velocity
by
$v = \tanh\kappa$, $E_i^\prime$ the energy of particle
$i$ and $\theta^\prime $ the angle
between the decaying particle, $A$, and the direction of the final
particle in the boosted frame.

We are interested in studying the density matrix of the final
neutrino in
the boosted frame in a decay $ P^+( q )\rightarrow \nu_l (
k_\nu,\lambda_\nu )
l^+ ( k_l, \lambda_l )$. Then it is easy to see that the density
matrix for the
neutrino transforms as a direct
 product of two helicity amplitudes if we average
over the polarization of the charged lepton.
Our boost will take us from the C.M. frame to the LAB frame, where
the meson has three-momentum $\mid \vec{q^{\prime}}\mid$ different
from zero.
Now  we can  get the density matrix of the $\nu_l$ in the
boosted frame, if we know it in C.M.:
\begin{equation}
\label{rot1/2}
\rho^{LAB} = d^{1/2} (\omega) \cdot \rho^{CM} \cdot d^{1/2\ T}
(\omega)
\end{equation}
where $\omega$ is the Wigner rotation asociated with the $\nu_l$.

As can be easily seen, in the C.M. frame we only have longitudinal
polarization. We apply the
transformation law (\ref{rot1/2}) to get the expression of the
density
matrix in the LAB frame.
\bea
\rho^{LAB} =\frac{1}{2}\left( \begin{array} {cc}
1 + P_{long} \cos \omega   &  P_{long} \sin \omega \\
 P_{long} \sin \omega  & 1 - P_{long} \cos \omega
\end{array} \right) \nn
 \eea
We can see that, in the new frame of reference, a new transverse
polarization has
appeared, $P_x =  P_{long} \sin \omega$, and the longitudinal
polarization has
been modified too: $P_z =  P_{long} \cos \omega$. This is a very
important
fact because, as we have
already pointed out, if $m_{\nu} = 0$, in every frame the neutrino
will
have only longitudinal polarization, and it will always be equal to
-1.
Moreover, notice that the effects on the transverse polarization are
linear in $m_{\nu}$ as shown in Eq. (\ref{wignercos}), so transverse
 polarization will be more
sensitive to small neutrino masses.

Let us try to analyse carefully the new situation in the boosted
frame.
Qualitatively, we have two very different situations:

i) the velocity of
the decaying particle in LAB is less than the velocity of $\nu_l$
in C.M.,

ii) the velocity of the decaying particle in LAB is
bigger than the velocity of $\nu_l$ in C.M.

In i), in LAB, we still have particles both in the forward and
backward hemispheres. Concerning the Wigner rotation we have an
important feature: for any value of $\theta^\prime$, $\cos \omega$
will
always be positive and consequently $\sin \omega$ will have a maximum
value strictly less than 1,
\bea
\left.
\sin \omega \right|_{max} =
\frac{m_\nu \sinh \kappa}{\mid \vec{k}_\nu \mid} =
\frac{m_\nu \mid \vec{q^\prime} \mid }{\mid \vec{k}_\nu \mid m_P}
\nn \eea

In ii), all the $\nu_l$ are now in the forward hemisphere in LAB.
{}From
kinematics, there is a maximum $\theta^\prime$ for  which
$\sin \omega = 1$. The value of  $\cos \omega $ can even change sign
for the
neutrinos that were in the backward hemisphere in the C.M. frame
(which appear forward in the LAB).
This means, for instance, that the neutrinos which in C.M. were
moving
opposite to the boost, have reversed the direction of its momenta
and they are now right-handed neutrinos.

As we can see, the effects of a boost on the neutrino polarization
are
really important.
We could ask  whether they also translate into an observable effect
of
the charged lepton polarization in the boosted frame.
 It would be much easier
to measure the charged lepton polarization
from the distribution of its
 decay products than the $\nu_l$ polarization,
and we could naively
expect them to be somehow related. In fact, in the C.M. frame
the charged lepton polarization is exactly opposite to the neutrino
polarization, but this is no longer true in a boosted frame,
due to the orbital angular momentum in this frame.
Because of this, we need to reconstruct completely the decay,
this means we have to know the tau and meson directions.
Experimentally, this is not an easy task, because one has two
neutrinos in each $D$ decay. Therefore, from the practical point
of view, unless we have an extremely good experimental
precision it is easier to deal with the neutrinos
directly. In this letter we are going to follow this avenue.
However, one should keep in mind that a complete reconstruction
of the decay would provide the information also from the tau.

Taking into account the available and foreseen experimental
facilities we choose to take profit of the existing and proposed
neutrino
beams \cite{E872}, \cite{jj} at FermiLab and LHC.
The Achiles' heel of this approach is that we have to measure
the neutrino polarization.
The transverse polarization is, a priori, the most sensitive one
to neutrino mass, but its measurement would
be rather complicated and would introduce additional masses.
Fortunately, to measure the longitudinal polarization will
be much easier.

If we boost our neutrinos  according to the situation ii),
a measurable fraction of the neutrinos will have reversed their
helicity.
Neglecting those interactions proportional to neutrino mass,
the gauge interactions are
left-handed, so that {\it{only}} those neutrinos with the
{\it {proper}} helicity will interact. This means that we
have to look for a {\bf { left-handed
 neutrino deficit}} in order to reveal these
effects.

Specifically,
the flux and the corresponding spectrum of boosted neutrinos
could be obtained from measurements on the $D$-mesons produced.
 The dynamical neutrino
mass effects discussed here provide different fractions of improper,
right-handed, neutrinos as a function of its mass and energy. So a
deficit of
left-handed neutrinos will be indeed a proof of neutrino
mass. Not only this, the amount of such a deficit will be a
 measure of the neutrino mass itself. The kind of
dissapearance will shed light into the character of the neutrino
type.
Clearly, the track left by a Dirac neutrino would be only
a depletion in the expected flux of active neutrinos due to the
sterile
nature of right-handed neutrinos.
However, a Majorana neutrino would bring in its right-handed
state new reactions associated with the charge conjugated
state.
Summing up, counting neutrino interations provides a
measure of its longitudinal polarization.

Following this strategy, we have to calculate the fraction of
left-handed neutrinos in the beam. To do so, we need the neutrino
spectrum which is given by $d\Gamma/dE_\nu = (16 \pi E_{q^{\prime}}
\mid\vec{q^\prime}\mid
)^{-1}$, at a fixed boost velocity, where $E_{q^{\prime}}$ is the
parent
meson energy in LAB.
As we need the spectrum of left-handed neutrinos, we just have to
multiply the energy spectrum by the fraction of left-handed neutrinos
$f(L)$ at a given energy
\bea
\frac{d\Gamma_L}{dE_\nu} =
\frac{d\Gamma}{dE_\nu}  f(L) \;\;\; \;
; \; \; \; \; f(L) = \frac{1 - P_z(E_\nu) }{2}
\nn \eea
where $P_z(E_\nu) = P_{long} \cos\omega $ is the neutrino
longitudinal polarization in the boosted frame at a given energy.

Figure 1 shows the fraction of left-handed neutrinos as a function
of the neutrino energy for a fixed boost (100 GeV) of the $D$ (a) and
$D_s$
(b) mesons
and a neutrino mass of 24 MeV.
As we can see, the effect is huge for the low energy neutrinos and it
dilutes
for increasing energy.

In Figure 2 we present the integrated spectrum of left neutrinos as a
function of the boost energy, normalized to the total flux of
neutrinos,
i.e, one minus the deficit.
It can be seen that above a boost of moderate energy around 20 GeV,
the
deficit becomes constant and we do not improve the integrated effect
by
increasing the boost energy. However, the energy of the improper
helicity  neutrinos increases with that of the boost. This is very
important if a cut in threshold energy is applied as discussed now.

According to Figure 2, for a boost above 20 GeV
 we can expect a depletion of a 7.5$\%$ for the
$D$ decay and a tau-neutrino of 24 MeV, if we integrate
the complete spectrum. This roughly quadruplicates what we can get
in C.M. measuring the tau polarization. When considering the
threshold
(2 GeV) in charged current reactions for detecting tau-neutrinos we
get a
sizable deficit of left-neutrinos for high velocity boosts: at 100
GeV
we find an integrated depletion of
4$\%$ above 2 GeV, and it increases with the boost towards the
asymptotic
7.5$\%$.
This is so because the fraction of wrong helicity neutrinos, which is
bigger in the
low energy part of the spectrum, moves upwards with the boost.
The deficit fraction would
increase for neutrinos below some energy cut, say 1 or 2 GeV,
reaching 36$\%$ or 23$\%$, respectively for a boost of 100 GeV.
For tau neutrinos, these low
energy neutrinos are only detectable through neutral
current scattering. Even through neutral current interactions the
distinction between Dirac or Majorana can be made \cite{s2}. Neutral
current events are dominated by $\nu_\mu$ events,
however these can be controlled and substracted by measuring the
charged current events at the same energy they give rise to.
The best strategy for the $\nu_\tau$-detection will depend on the
actual
characteristics of neutrino beams \cite{E872}, \cite{jj}. For boosts
high enough, charged current detection is simpler and the number of
counts
increases linearly with the energy. In particular, for Majorana
neutrinos,
the presence of a wrong sign charged lepton constitutes a clean
signature.

The situation is not so promising for the $D_s$ decay. In this case,
we get a deficit of
approximately 2.5$\%$ when we integrate the complete spectrum.
Putting an upper cut in the energy, in this case, gives a deficit
of 22$\%$ (1 GeV) or 13$\%$ (2 GeV), again for a boost
of 100 GeV.

To conclude:
in this work we have studied the viability of a dynamical
determination
of the tau
neutrino mass using branching ratios and longitudinal polarizations,
instead of the direct kinematical approach.
In light of kinematics, the most favourable situation is the
$D$-meson
 decay which allows us to get effects of at most 4$\%$ in the C.M.
longitudinal polarization.
The situation becomes more promising if we take advantage of the fact
that, while the branching ratio is a Lorentz-invariant quantity, the
longitudinal polarization is not. Under appropriate conditions, the
polarization can be dramatically modified by a boost. We have
discussed
possible strategies to observe the enhanced deficit of
left-neutrinos.

\begin{center}
{\bf ACKNOWLEDGMENTS}
\end{center}

We are indebted to S. Bilenky, F.J.Botella, M.C.Gonzalez-Garcia,
J.J.G\'omez-Cadenas and
A.Pich for stimulating discussions and provocative
questions.
G.B. acknowledges the Spanish Ministry of
Foreign Affairs for a MUTIS fellowship, whereas O.V. does it
to the Generalitat Valenciana for a research fellowship.
This work was supported by CICYT, under grant AEN 96/1718.

\pagebreak

{\large{\bf { Table I}}}

\begin{table}[htb]
\begin{center}
\begin{tabular}{|l||l|l|l|l|l|}
\hline
$m_{\nu_\tau} $  ( MeV) & 0 & 6 & 12 & 18 & 24 \\\hline
\hline
$ -P_{long}$ & 1 & .9975 & .9901 & .9778 & .9603 \\\hline
\hline
$ \frac{\Gamma (D \rightarrow \tau \; \nu_\tau)}{\Gamma (D
\rightarrow
\mu \; \nu_\mu)}$ & 2.640 & 2.636 & 2.622 & 2.598 & 2.563 \\\hline
\end{tabular}
\end{center}
\end{table}

\vspace{.3cm}

{\large{\bf { Table II}}}

\begin{table}[htb]
\begin{center}
\begin{tabular}{|l||l|l|l|l|l|}
\hline
$m_{\nu_\tau} $  ( MeV) & 0 & 6 & 12 & 18 & 24 \\\hline
\hline
$-P_{long}$ & 1 & .9993 & .9973 & .9940 & .9893 \\\hline
\hline
$ \frac{\Gamma (D_s \rightarrow \tau \; \nu_\tau)}{\Gamma (D_s
\rightarrow
\mu \; \nu_\mu)}$ & 9.739 & 9.736 & 9.726 & 9.710 & 9.687 \\\hline
\end{tabular}
\end{center}
\end{table}

\vspace{.3cm}

{\large{\bf { Table captions}}}

\vspace{.3cm}

{\bf {Table I : }} CM frame observables for
the decay $D \rightarrow \tau \; \nu_\tau$. The first
row shows the neutrino-tau mass effects in the
longitudinal polarization, while the second row shows
the branching
ratio between the decay rates of the $D$ meson to tau and muon
for different masses of the tau-neutrino (the muon-neutrino
is taken to be massless).

\vspace{.3cm}

{\bf {Table II : }}  As Table I, but for the decay
 $D_s \rightarrow \tau \; \nu_\tau$.

\vspace{.3cm}

{\large{\bf { Figure captions}}}

\vspace{.3cm}

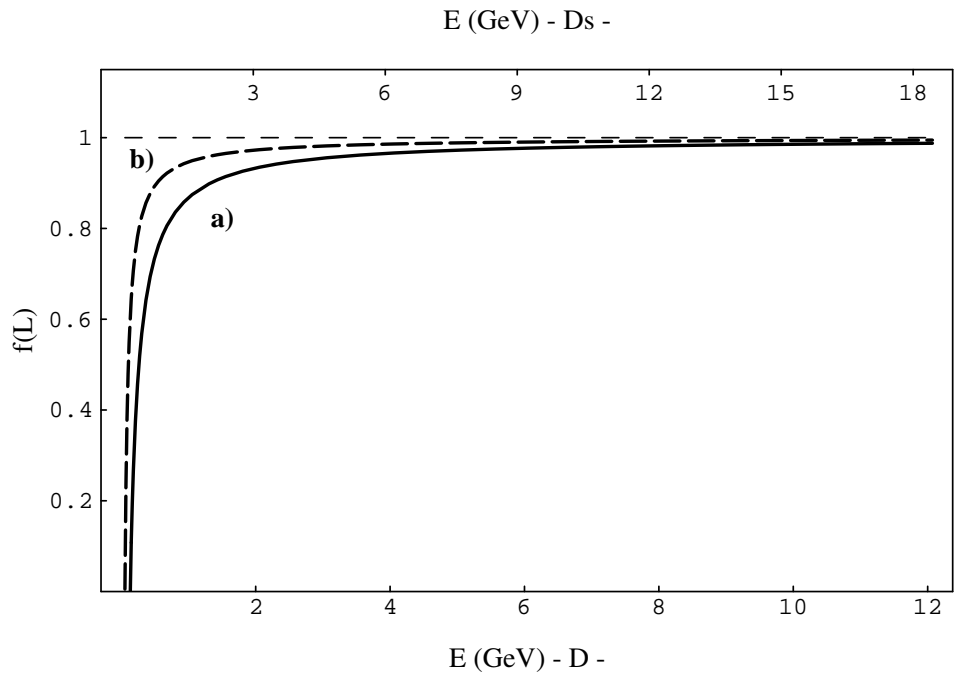
{\bf {Figure 1 : }}Fraction of left-handed neutrinos as a function
of their energy for a neutrino mass of 24 MeV and a boost of 100
GeV in the $D$ (a) and $D_s$ (b) decays.

\vspace{.3cm}
{\bf {Figure 2 : }}Left-handed fraction of the total neutrino flux,
$R =
\Gamma_L /\Gamma$,
as a function of the parent meson boost for $D$ (solid line) and
$D_s$ (dashed line) decays.

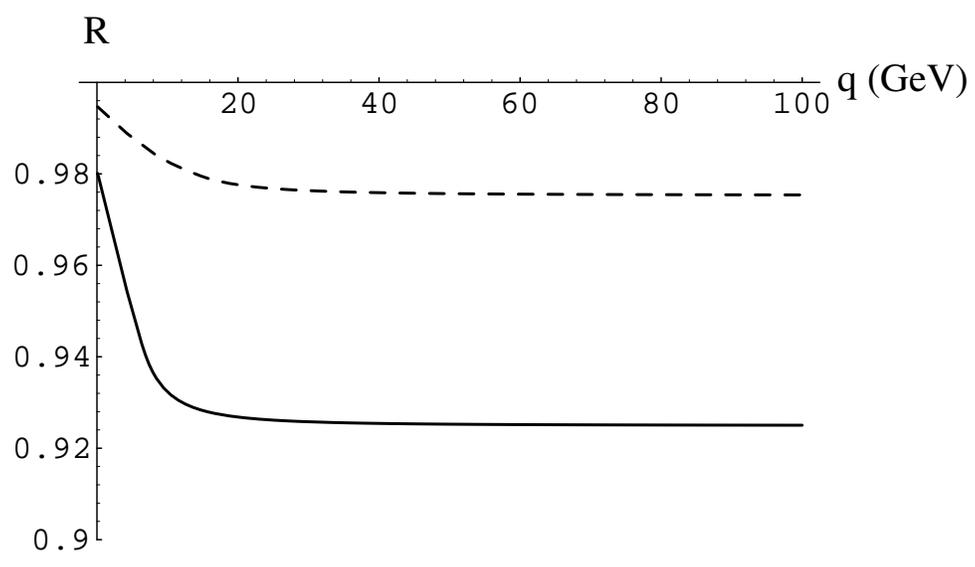
\end{document}